# Medipix3 – Demonstration and understanding of near ideal detector performance for 60 & 80 keV electrons.

J.A. Mir[a,*], R. Clough[a], R. MacInnes[c], C. Gough[c], R. Plackett[b], I. Shipsey[b], H. Sawada[a], I. MacLaren[c], R. Ballabriga[d], D. Maneuski[c], V. O'Shea[c], D. McGrouther[c], A.I. Kirkland[a]

a University of Oxford, Department of Materials, Parks Road, Oxford OX1 3PH, United Kingdom

b University of Oxford, Department of Physics, Parks Road, Oxford OX1 3PH, United Kingdom

c University of Glasgow, School of Physics and Astronomy, Glasgow G12 8QQ, United Kingdom

d CERN,1211 Geneva 23, Switzerland

* Corresponding author. E-mail address: jamil.mir@materials.ox.ac.uk

## HIGHLIGHTS

- Near-ideal *DQE* & *MTF* detector performance obtained using a transmission electron microscope at 60 and 80 keV
- Demonstration that Charge Summing Mode simultaneously maximises both *DQE* & *MTF*
- Understanding of detector response through analysis of single electron events
- Demonstration of 24-bit depth high dynamic range and charge summing mode imaging

# ABSTRACT

In our article we report first quantitative measurements of imaging performance for the current generation of hybrid pixel detector, Medipix3, as direct electron detector. Utilising beam energies of 60 & 80 keV, measurements of modulation transfer function (*MTF*) and detective quantum efficiency (*DQE*) have revealed that, in single pixel mode (SPM), energy threshold values can be chosen to maximize either the *MTF* or *DQE*, obtaining values near to, or even exceeding, those for an ideal detector. We have demonstrated that the Medipix3's charge summing mode (CSM) can deliver simultaneous, near ideal values of both *MTF* and *DQE*. To understand direct detection performance further we have characterized the detector response to single electron events, building an empirical model which can predict detector *MTF* and *DQE* performance based on energy threshold. Exemplifying our findings we demonstrate the Medipix3's imaging performance, recording a fully exposed electron diffraction pattern at 24-bit depth and images in SPM and CSM modes. Taken together our findings highlight that for transmission electron microscopy performed at low energies (energies <100 keV) thick hybrid pixel detectors provide an advantageous and alternative architecture for direct electron imaging.

# KEYWORDS

# 1. INTRODUCTION

Direct electron detection can be accomplished by using the conventional film or by using solid-state detection technology such as the Monolithic Active Pixel Sensors [1] or a variant hybrid pixel detector technology [2,3] such as the Medipix3-based detectors. In hybrid pixel detectors, the detector consists of a semiconductor absorber layer connected to a readout ASIC that processes the signal induced in the sensor. It is well known that thin hybrid silicon detectors with small pixels are well suited for obtaining optimum images at higher TEM voltages, typically 200 keV and beyond. In this study, we demonstrate, using the Medipix3, that a thick silicon hybrid with coarse pixel geometry is ideal for low voltage TEM imaging up to 80 keV. In this voltage regime, the *MTF* is almost invariant and yields high *DQE*s.

The Medipix3 detector was designed at CERN in the framework of the Medipix3 collaboration for photon and particle detection using the commercial 0.13 μm CMOS technology and measures 15.88x14.1 mm$^2$. The matrix consists of 256x256 pixels at 55 μm pitch. The readout chip was connected to a 300μm thick Silicon layer. Each pixel contains analogue circuitry consisting of a charge sensitive preamplifier, a semi Gaussian shaper and a two discriminators that control the lower and upper threshold levels. Each discriminator has a 5-bit Digital to Analogue Converter (DAC) to reduce the threshold dispersion caused by mismatch in the transistors. In the Single Pixel Mode (SPM), pixels only register a count if the induced energy exceeds the preset lower threshold energy value, TH0. Each pixel also contains two configurable depth registers which can also function as counters enabling a continuous Read-Write capability whereby one register acts as a counter whilst the other shifts the data out.

When compared with the Medipix2 detector [4,5], the Medipix3 design contains an additional functionality called Charge Summing Mode (CSM) designed to mitigate the effects of charge sharing. Charge sharing occurs when the charge produced, by an incident electron or photon undergoes lateral dispersion due to electron scattering or diffusion and is spread across several pixels leading to degradation in both energy and spatial resolution. The situation for an electron beam entering the detector is much worse than for photons, as unlike the localized photon absorption, electrons lose energy sporadically through inelastic scattering events distributed over micrometer scale distances (for beam energies of the order of tens of keV). The Medipix3 detector has been designed to minimize the effect of charge sharing by allowing a specific mode whereby charge is deposited in clusters of immediate neighbouring pixels is summed, at pixel corners, followed by allocating the reconstructed charge to the individual pixel with the highest collected charge. This is accomplished in several steps. For example, if charge created from the initial event encompasses four pixels then the individual pixel charges are compared to a selectable energy threshold TH0. The digital circuitry on the pixels processes the pulses to identify the pixel with the largest charge and inhibits the pixels with lower signal. In parallel the charge is reconstructed in analog summing circuits located effectively at the corners of each pixel and compared to an energy threshold, TH1. The pixel with the highest local charge increases its counter if the reconstructed charge on at least one of its adjacent summing nodes is above TH1 [6,7].

We have used a single chip Medipix3 detector to investigate its performance for TEM at 60 and 80 keV. The basic metrics for quantifying the detector performance are the Modulation Transfer Function (*MTF*) and the Detector Quantum Efficiency (*DQE*). The *MTF* is the ratio of output to input modulation as a function of spatial frequency and effectively describes how the detection system attenuates the amplitudes of an infinite sinusoidal series. In the present work, we used the established knife edge method [8] to derive the *MTF* as well as a new technique to calculate the Point Spread Function (*PSF*) directly from short exposure flat-field images capturing single electron events.

In the case of the knife-edge method, a Line Spread Function (*LSF*) was initially obtained by differentiating the experimentally obtained edge profile. The modulus of the Fast Fourier Transform (FFT) of the *LSF* yielded the *MTF*. The *DQE* is the ratio of square of the output to the square of the input Signal-to-Noise (*SNR*):

$$DQE(f) = \frac{[SNR_{out}(f)]^2}{[SNR_{in}(f)]^2} \quad (1)$$

where *f* is the spatial frequency. The *DQE* can be calculated with the knowledge of *MTF* and the Noise Power Spectrum (*NPS*) as follows [5]:

$$DQE(f) = \frac{c^2 MTF^2}{n(NPS)} \quad (2)$$

where *c* represents the number of counts in the output image and n is the electron input (dose). In order to calculate the *DQE*, a knowledge of the *MTF*, *NPS* and the gain factor, *g*, defined as the ratio *c*/*n* for a given operational voltage was required.

The *NPS* was calculated from the FFT of the flatfield images. Both the *MTF* and the *DQE* were evaluated in the spatial frequency range of 0 to 0.5 pixel$^{-1}$ where the upper limit represents the Nyquist frequency beyond which aliasing occurs. In the present case, this limit corresponds to 9.1 lp/mm. As noted in [5], it is difficult to calculate the *NPS* and hence the *DQE* at lower spatial frequencies accurately. The observed variance in a flat-field image results in underestimation of the true noise per pixel as the charge produced by an incident electron is seldom confined to a single pixel. We have carried out a similar analysis to [5] using equation (1) to calculate the DQE at the zero spatial frequency *DQE* (0), in section 2 below.

The *MTF* of the detector arises fundamentally due to the manner in which electrons deposit their energy in the silicon sensor slab and by which the resultant electron-hole pairs diffuse under bias toward the ASIC bump bonds. Assuming the energy required to produce a single electron-hole pair is 3.6 eV in silicon, a single primary electron at 60 keV can produce over 16,000 electron-hole pairs. Owing to the excellent SNR provided by Medipix3, we have been able to perform analysis of single electron events during short shutter exposures, with duration in the range 1-10μs, revealing single and multi-pixel clusters. Characterisation of cluster area and detector response was performed as a function of threshold energy and synthetic PSFs calculated. *MTF*s were produced by direct Fourier transformation of the PSF.

## 2. EXPERIMENTAL

The Medipix3 detector was mounted on the JEOL ARM200cF TEM/STEM [9] n a custom constructed mount into the 35 mm camera port located above the viewing screen. This mount included a vacuum-tight feedthrough for a 68 way electrical connector to allow readout. Operation and high speed data readout of the detector was via MERLIN hardware/software produced by Quantum Detectors [10,11]. The *MTF* and *DQE* data was taken for primary electron beam energies of 60 and 80 keV using SPM mode. For each primary electron energy, the *MTF* data was taken by recording images of a 2 mm thick aluminium knife edge inclined by 10 degrees with respect to the pixel readout columns. Having set the exposure time to 10 ms, 32 repeated images were acquired across the full range of Medipix3 energy threshold values in the SPM mode. The *MTF* data acquisition procedure was then repeated using CSM by holding TH0 DAC at a fixed energy and scanning the high threshold (TH1) DAC across the full range of energy values.

For both primary electron energies (60 and 80 kV), a set of 32 flat-field images were acquired as a function of threshold energy values for calculating the *NPS*, and ultimately the *DQE* using equation (2). This was accomplished with 10 ms and 1 µs exposure times using SPM and CSM modes. Energy calibration of the energy thresholds was performed by taking into account certain criteria. For both SPM & CSM, modes, at the incident beam energy, the total integrated counts must go to zero. In addition, for SPM mode, additional calibration points were available by identifying the threshold DAC values where electrons resulted in only single pixel hits, i.e. at the half the incident beam energy [5]. The gain factors, *g*, for the specified primary electron energies were measured by projecting the full electron beam diameter within the detector perimeter and recording 32 images as a function of a number of different threshold energy values. For each voltage used, the beam current was measured focusing the beam within the beam current measuring region located at the top of the small viewing screen and measuring current with a Keithley 485 Picoammeter. The following equation summarises the dependence of *g* on the measured current, exposure time and the pixel sum values:

$$g = \frac{Current \; x \; Exposure \; time}{Pixel \; Sum \; x \; e} \tag{3}$$

## 3. DETERMINATION OF *MTF* & *DQE*

*DQE(0)* was measured using the method described in [5,12] where the noise $(N_x)^2$ is measured in *x* by *x* binned images with increasing *x*. Specifically 32 flat-field images were analysed and each image had the previous image in the series subtracted to produce 31 images of uniform illumination with a mean pixel value. The noise per pixel is found by plotting $(N_x)^2/x^2$ as a function of $x$ and recording the plateau. Figure 1 shows an noise evaluation for calculating *DQE*(0) using 60 keV flat-field SPM images for a number of different threshold DAC values. As expected the noise per pixel reduces with increasing threshold DAC values, since the variance reduces with decreasing effective pixel size. After extracting the noise per pixel values at plateau, the *DQE*(0) for a for a given DAC value was calculated using equation (4) below.

$$DQE(0) = \frac{c^2}{0.5\left(\frac{N_x^2}{x^2}\right)/n} \quad (4)$$

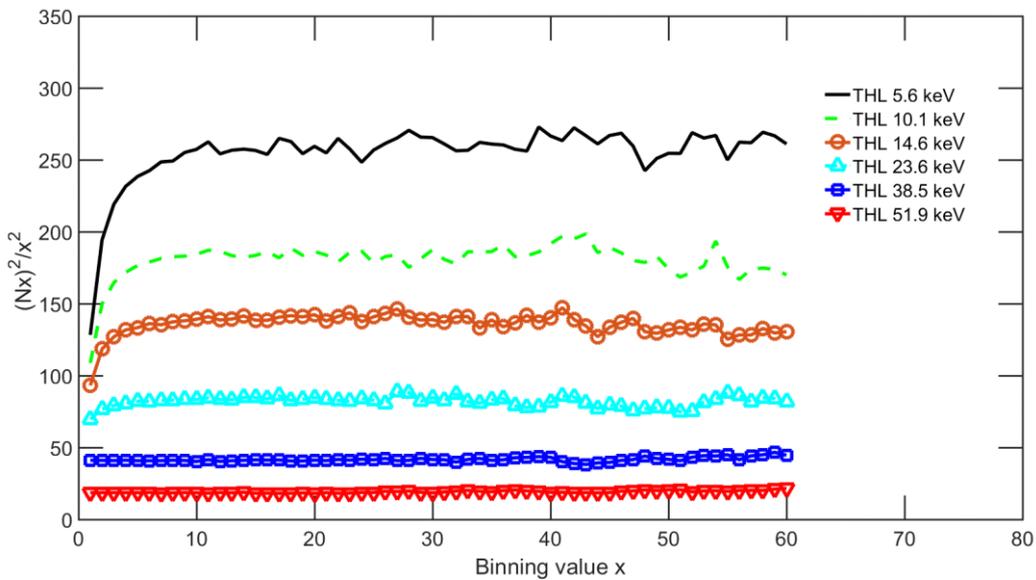

*Figure 1: Variation of the noise, $(N_x)^2/x^2$, as a function of x-fold binning at 60 keV for various TH0 thresholds.*

Figures 2 (a-b) show the variation of *MTF* as a function of spatial frequency at 60 and 80 keV, respectively, with single pixel mode for various TH0 energy thresholds. These figures also show the theoretical *MTF* response of an ideal detector given by the function sinc(π$f$/2). At the highest lower threshold (TH0) values the *MTF* for this counting detector in single pixel mode is better than the theoretical maximum due to the reduction in the effective pixel size [4]. However, the *DQE* at such high TH0 DAC values in single pixel mode is significantly reduced as seen in Figures 3 (a-b) and 4, seeing as many real electron events are now not counted as the charge is deposited in more than one pixel and therefore falls below the threshold for detection. Consequently, there is a balance to be made between optimizing *DQE* and *MTF*, depending on the exact requirements in the given application. Figures 3 (a-b) also show the theoretical *DQE* response of an ideal detector as given by the function sinc$^2$(π$f$/2). In addition, the *DQE*(0) values shown in these figures were calculated independently from the analysis of the flat-field images using equation (4). Figure 5 shows the variation of *DQE*(0) as a function of threshold DAC values.

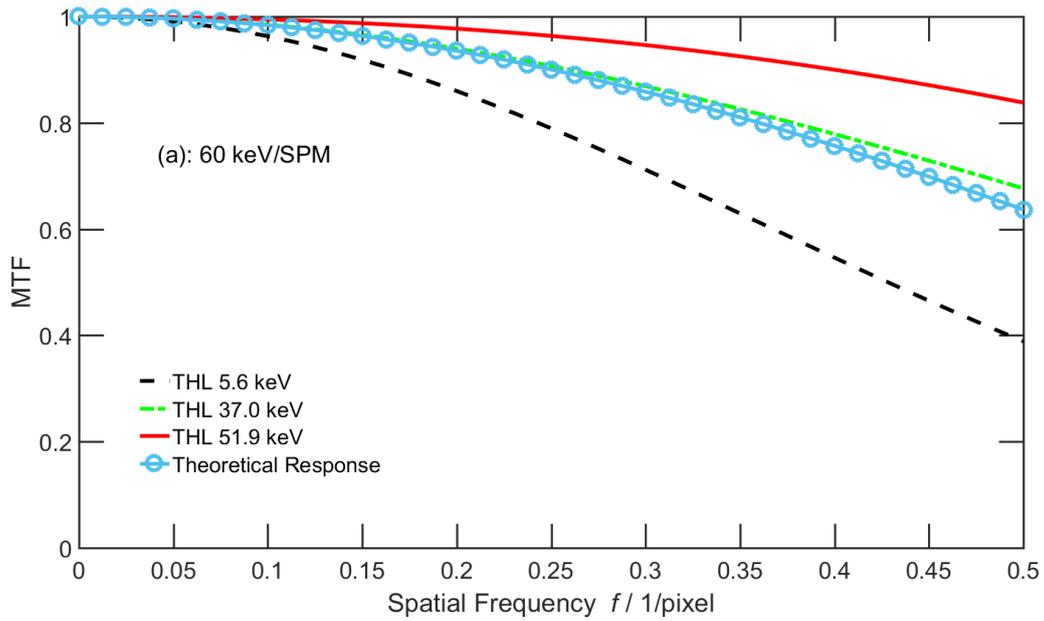

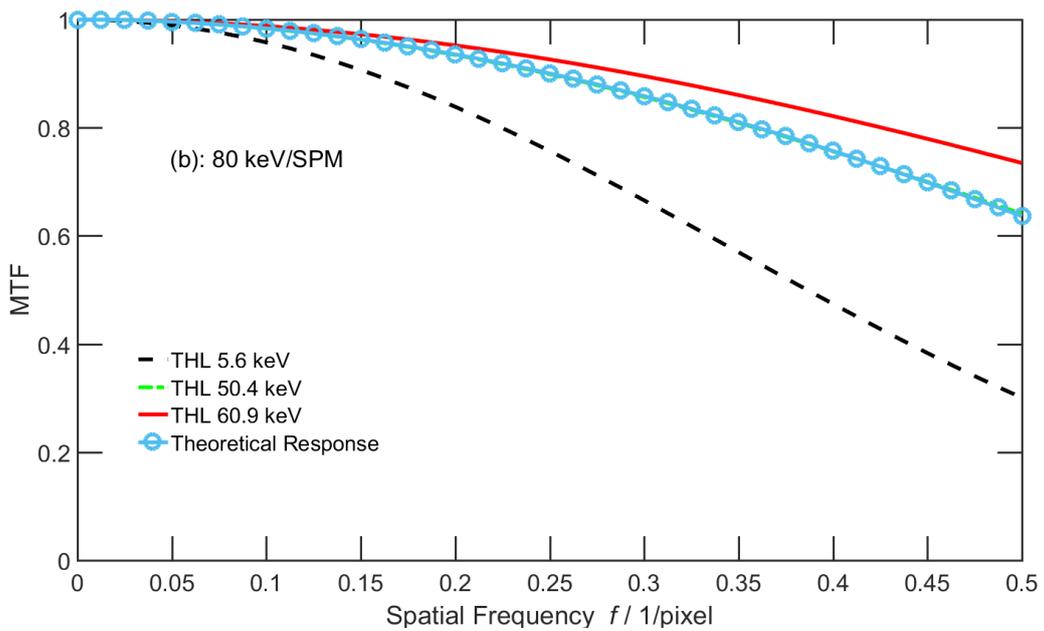

*Figure 2(a-b): MTF as a function of the spatial frequency at 60 and 80 keV with Single Pixel Mode (SPM) for various TH0 DAC values. The theoretical response of an ideal detector is illustrated by the curve with circular markers.*

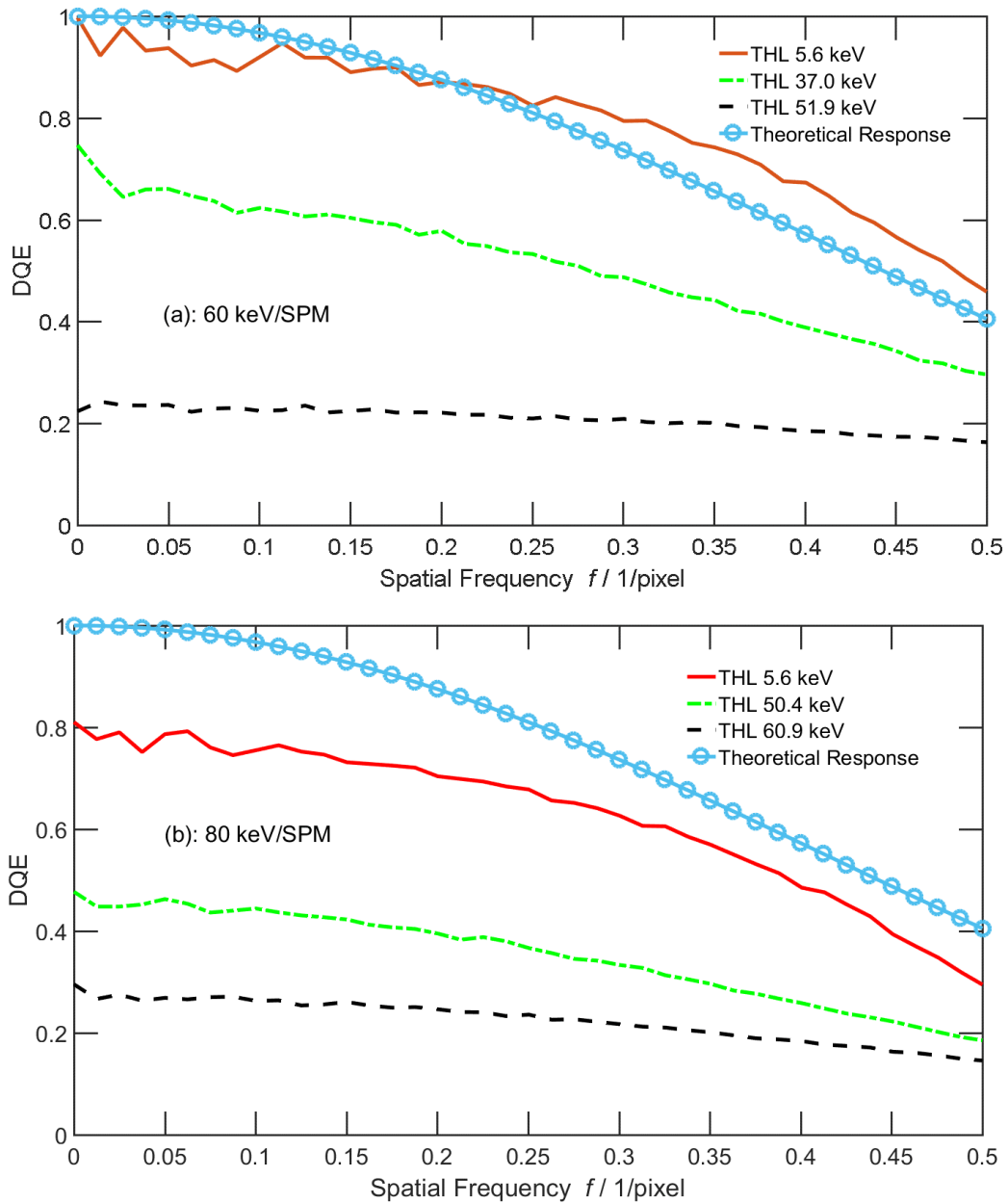

*Figure 3(a-b): DQE as a function of the spatial frequency at 60 and 80 keV with Single Pixel Mode (SPM) for various TH0 DAC values. The theoretical response of an ideal detector is illustrated by the curve with circular markers.*

The degradation of *MTF* seen in Figures 2 (a-b) with increasing primary electron energy is consistent with earlier work [5] and by Monte Carlo simulations using CASINO[13] which show that the lateral charge spread (95%) at 60 keV is approximately 25 µm and increases to approximately 42 µm at 80 keV when entering a 300 µm thick silicon substrate.

At higher energies, electron dispersion increases leading to pixels being triggered at further distances from the initial incidence. The reduction in *MTF* with electron energy impacts *DQE* proportionally (see equation 2) as shown in figures 3 (a-b) and figure 4 since electron scattering cross sections decrease with increasing primary electron energy. The reduction of *DQE* with increasing threshold DAC values is attributed to smaller proportion of electrons that exceed the thresholds being detected.

Figure 5 shows the variation of *DQE(*0) as a function of TH0 threshold for 60 and 80 keV electrons. It is evident that the slope of the *DQE*(0) curve changes when the TH0 threshold is set at half the primary electron energy. Above this point, the noise power spectrum is constant and only single pixels are triggered. Conversely, several pixels may be triggered by a single electron if the threshold is set below this point.

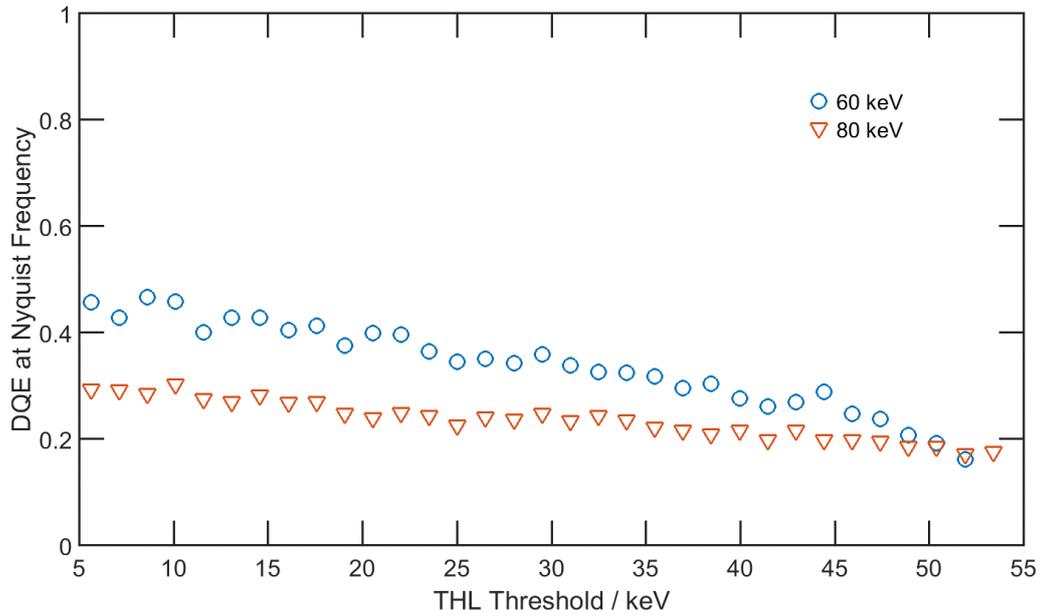

*Figure 4: Variation of DQE at the Nyquist frequency as a function of TH0 threshold values using SPM.*

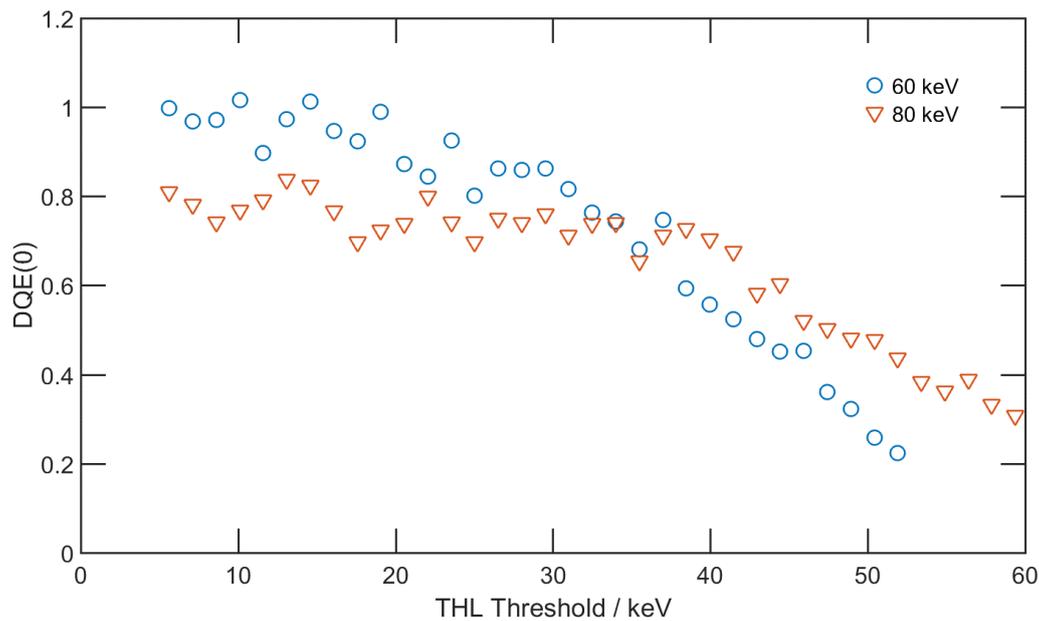

*Figure 5: Variation of DQE(0) as a function of TH0 threshold values using SPM.*

One of the major additional design advantages incorporated in Medipix3 detector over its predecessors is the charge summing mode where it collects charge from the neighbouring pixels and triggers the pixel with the highest fractional charge. In principle, the use of CSM should allow the achievement of high *MTF* performance but without having to set high values energy threshold to reject electrons which deposit energy across many pixels, as in SPM. Thus, in CSM, with low energy threshold values (with TH0 being the single pixel "arbitrated[1]" threshold and TH1 being the summed charge threshold), almost all detected electrons will be retained, maximizing simultaneously the *DQE*. This should allow very high efficiency imaging whilst preserving maximal detail in the images, which would have predictable benefits for imaging beam dose sensitive materials. Figures 6 and 7 record the *MTF* and *DQE* performance for CSM. From figure 6(a), at 60 keV, it can be seen that CSM provides almost ideal detector performance with little variation between the three TH1 energy thresholds plotted. In figure 6(b), at 80 keV, the *MTF* performance has reduced further with respect to the ideal but still maintains what would be regarded as high values across the spatial frequency range. Figure 7(a) & (b) shows that the high *MTF* values are matched by high *DQE* performance across the spatial frequency range, with only a weak spread for the TH1 energy threshold values plotted. In figure 7(a), at 60keV, the *DQE* values occupy a narrow band, being at most 0.2 lower the theoretical response of an ideal detector. At 80keV, in figure 7(b), the three thresholds plotted occupy a similar narrow band with values at most, 0.35, below the ideal response.

Comparison of the *MTF* performance between SPM and CSM modes is facilitated by figure 8 which plots the *MTF* at the Nyquist frequency as a function of threshold for 60 and 80 keV electrons for both SPM and CSM cases. It can be seen that *MTF* enhancement has been obtained at the lowest energy thresholds using CSM at 60 and 80 keV. In particular, an *MTF* at Nyquist frequency value around 0.6

---

[1] Single pixel arbitrated means that for a given hit, the counter associated to the threshold TH0 increases if the signal is above TH0 and the signal is the largest in its neighbourhood (This is different from the traditional Single Pixel Mode whereby the counter increases only if the signal is above TH0)

is achieved for 60 keV electrons (ideal detector *MTF* at Nyquist frequency = 0.64) when the energy threshold is set to its lowest value, 19.7keV, just above the Medipix3 chip's thermal electronic noise floor. *DQE* performance can be compared across the two modes by referring to figures 3 and 7. In SPM mode, figure 3, the *DQE* exhibited a strong inverse dependence on TH0 energy threshold. That is, low threshold energy values yielded highest DQEs but poorest *MTF* (as highlighted in figure 8). In CSM mode, figures 7(a) and (b), there is no longer a strong dependence on energy threshold and the *DQE* performance is similar, but slightly lower than the lowest SPM energy thresholds (TH0= 4.5 keV) in figure 3.

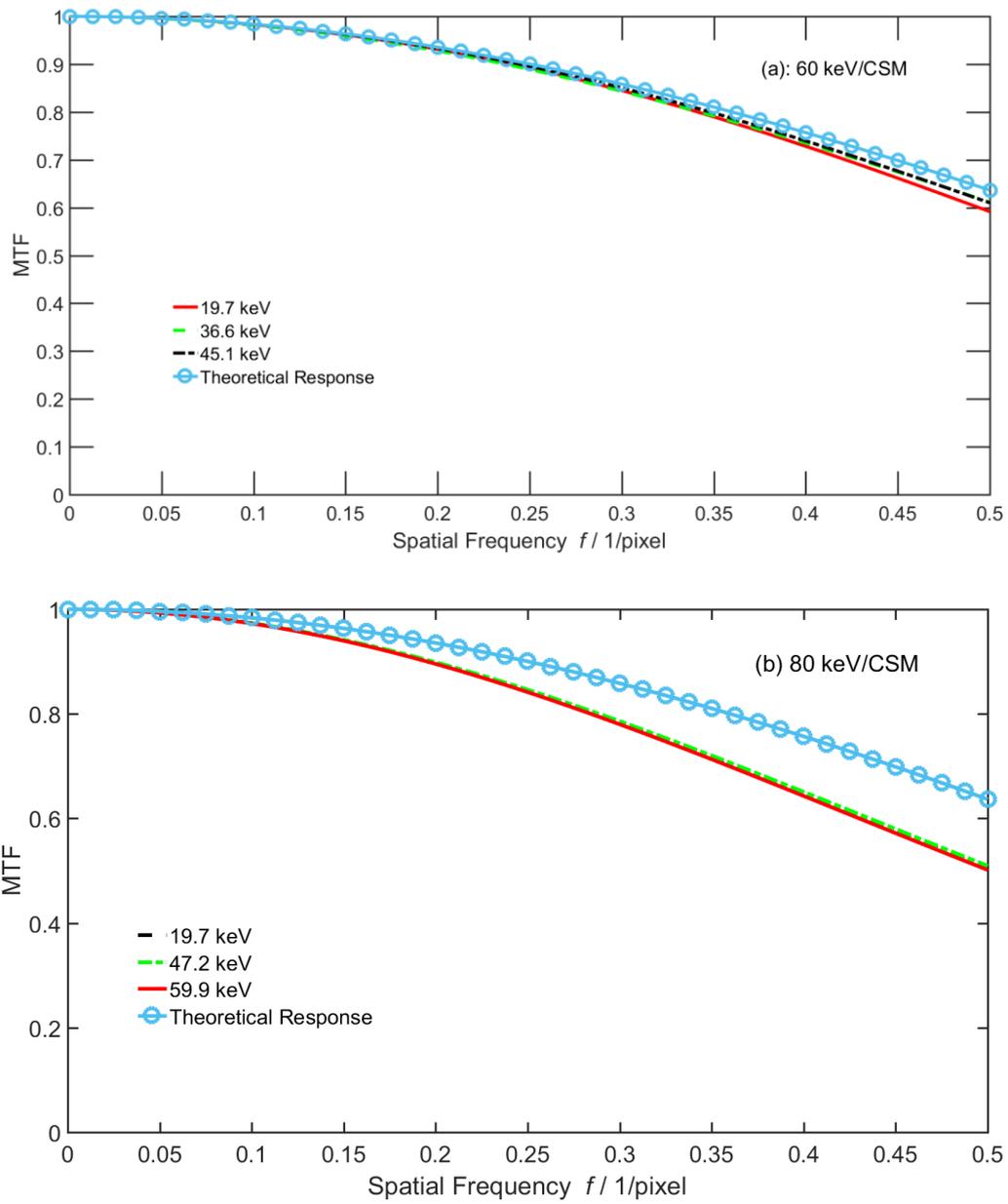

*Figure 6(a-b): MTF as a function of the spatial frequency at 60 and 80 keV with Charge Summing Mode (CSM) for various TH0 DAC values. The theoretical response of an ideal detector is illustrated by the curve with circular markers.*

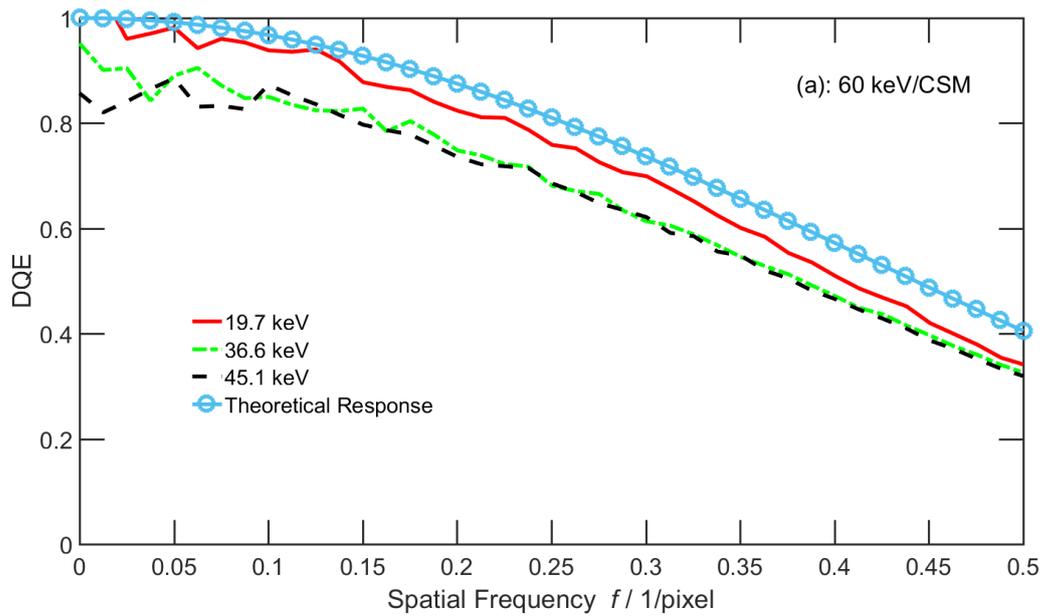

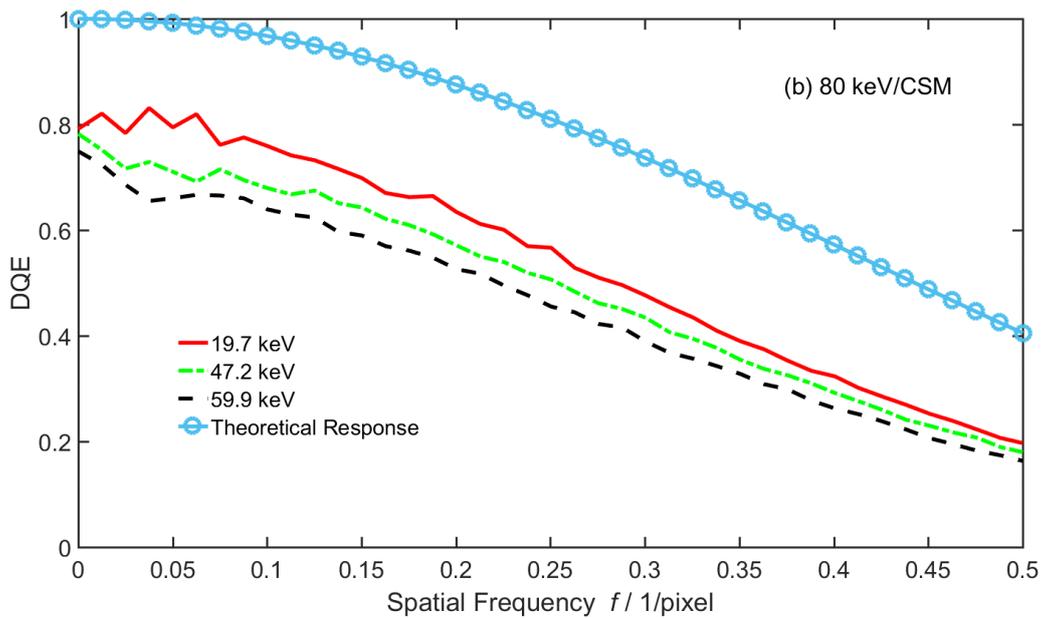

*Figure 7(a-b): DQE as a function of the spatial frequency at 60 and 80 keV with Charge Summing Mode (CSM) for various TH0 DAC values. The theoretical response of an ideal detector is illustrated by the curve with circular markers.*

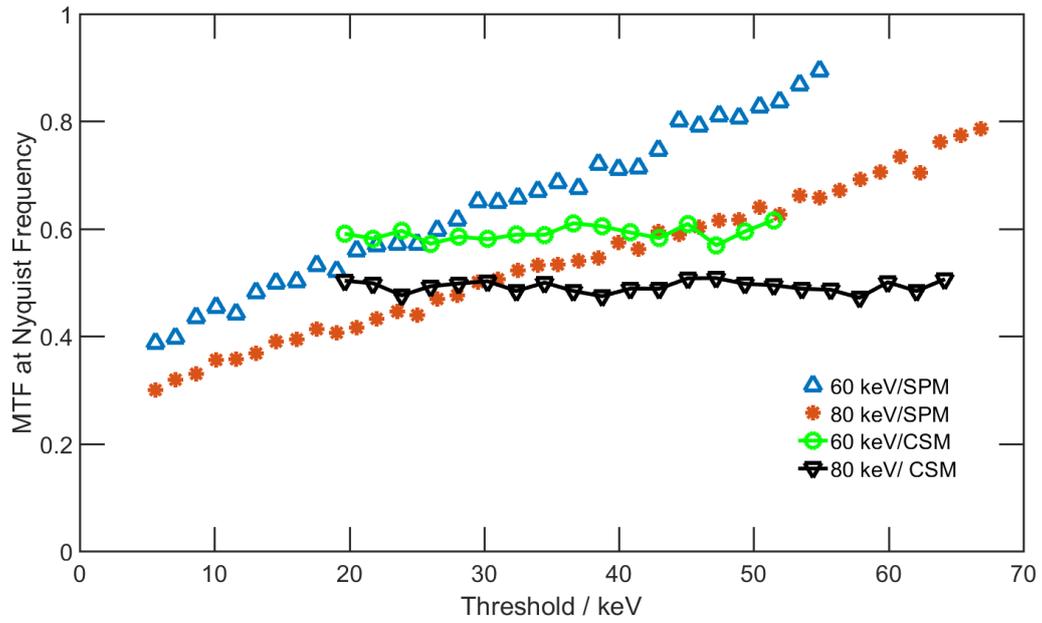

*Figure 8: A plot showing MTF at the Nyquist frequency using SPM and CSM at 60 and 80 keV. The TH0 in CSM at 60 and 80 keV were set at 4.5 keV for both.*

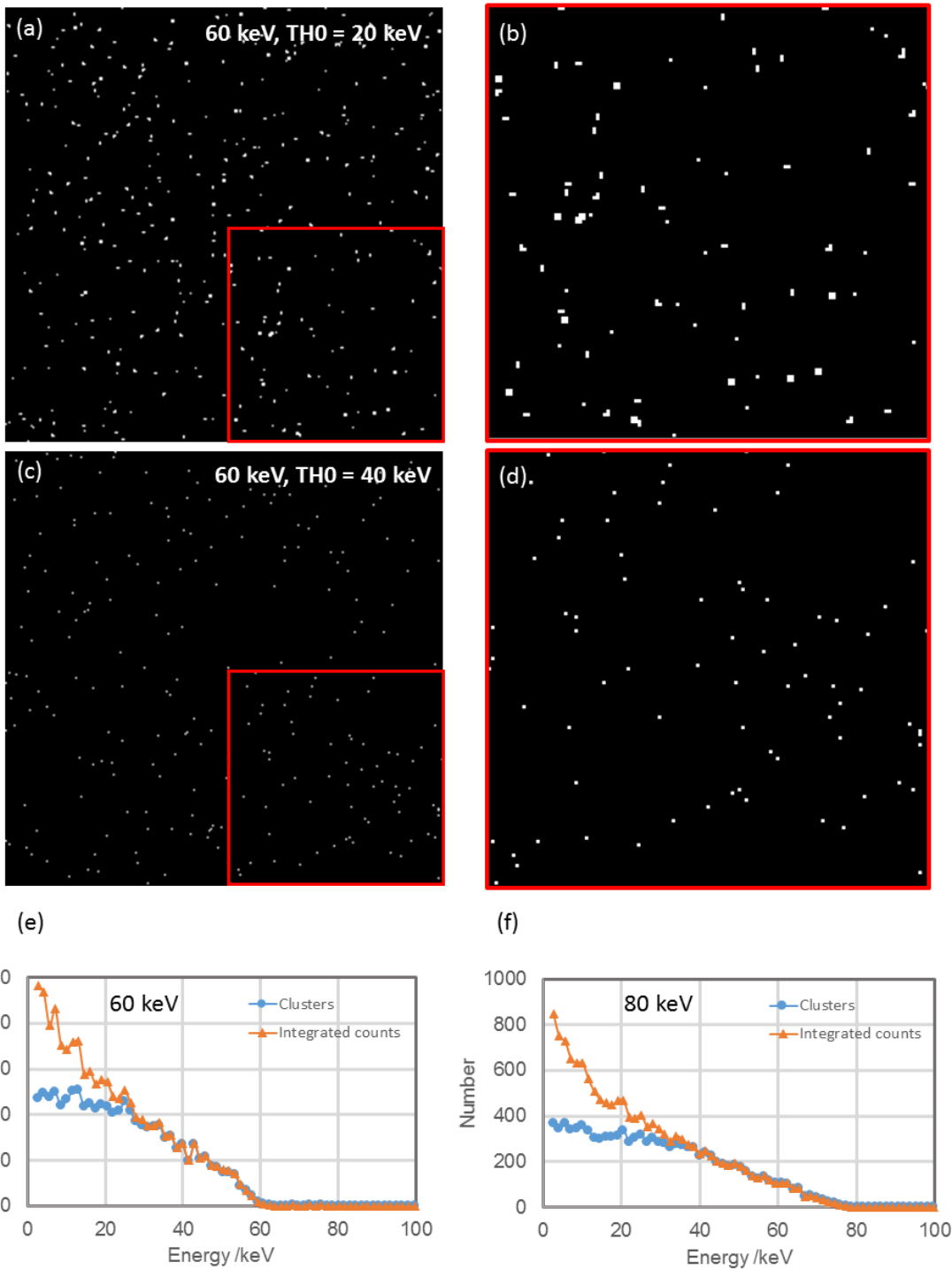

*Figure 9. (a) Single 60keV electron events with threshold energy, TH0=20keV. (b) Enlarged section from (a) showing multi-pixel clusters triggered due to charge spreading. (c) Single 60keV electron events with TH0=40keV. (d) Enlarged region from (c) showing only single pixels are triggered. (e) & (f) plots showing variation of integrated image counts and number of clusters counted vs TH0 energy for 60 keV and 80 keV electrons respectively.*

# 4. INVESTIGATION OF SINGLE ELECTRON EVENTS

In order to understand the behaviour of the detector, it's response to single electron events was studied through acquiring flat field images with an exposure time t = 10μs. Figures 9(a) and (b) shows the response of the detector to single $E_0$ = 60keV electrons when the threshold energy, TH0 is set to 20 keV (=$E_0$/3). Figure 9(b) reveals clearly that portions of the charge generated when single electrons impinge on the sensor slab can be deposited into pixels neighbouring the struck pixel, creating multi-pixel clusters. In figure 9(b) clusters can be seen to have areas of 1, 2, 3, and 4 pixels. By increasing the threshold energy TH0 to 40keV (=2$E_0$/3), figures 9(c) and (d) show that only single pixel clusters are obtained but that the overall number of clusters can be adjudged to have decreased. This is in accordance with previous measurements by McMullan *et al.* [4] of the Medipix2 detector where they showed that for TH0>$E_0$/2 only single pixel hits are obtained. The variation of the number of clusters counted with respect to threshold energy is plotted in figure 9(e). Also plotted in figure 9(e) are the integrated counts obtained from simple summation of all pixel values. Since the average separation of clusters is relatively large for the combination of beam current (105pA) and shutter time used, the integrated counts can be thought of as resulting from the number of clusters counted multiplied by their area in pixels. As such, when TH0 > $E_0$/2 = 30keV in figure 9(e) the two lines become superposed as only single pixel (unity area) clusters are obtained. Overall both the number of clusters and integrated intensity decrease (but not monotonically) from the maximum value at TH0 = 0 keV to zero at TH0 = $E_0$ = 60 keV. Similar behavior is obtained for $E_0$ = 80 keV electrons in figure 7(f) where a slightly greater number of clusters were counted at TH0 = 0 keV due to higher beam current (120pA).

The plots in figures 9(e) and 9(f) can be decomposed and empirically fitted in order to understand energy deposition and charge sharing in the detector. Naming the variation of the integrated counts with respect to TH0 as Σ(E), and the cluster counts variation with TH0 as N(E), a simple relation can be formed where Σ(E) = C(E) × N(E). The introduced function C(E) describes the variation of the average area of clusters with respect to TH0. For the data plotted in figures 9(e) and (f) the function C(E) is described by the expression: $C(E) = 1 + a\, e^{-E\delta}$ where *a* and *δ* are the fitted parameters whose values depend on the electron beam kinetic energy E0. C(E) fits are plotted for $E_0$ = 60 and 80 keV electrons in figure 10(a) where the data points result from dividing integrated counts data, Σ(E), by cluster counts data, N(E). From figure 8(a) it can be seen that there is relatively good agreement between the data and the fitted function. Examining only the cluster count data, N(E), in figure 10(b), it can be seen that a fit of the form:

$$N(E) = N_0\, n(E) = N_0\, erfc\left[\frac{\frac{1}{\sqrt{2}}(E - E_m)}{w}\right] \quad (5)$$

where *erfc* is the complementary error function and $N_0$, $E_m$ and *w* are the fitted parameters, describes the N(E) cluster count data relatively well. By forming the product of the C(E) and N(E) fits in figures 10(a) and (b) the Σ(E) data can then be well described as shown in figure 10(c).

Like the C(E) function, the N(E)=$N_0$ n(E) function also describes a variation in effective pixel area. At TH0 = $E_0$/2, when C(E) = 1 pixel area, Σ(E) becomes equal to N(E). As the value of TH0 is increased beyond $E_0$/2, the number of pixels registering hits decreases. This is because energies greater than the threshold value can only be received if pixels are struck in a zone located around the pixel centre with radius less than the pixel half-width. The zone radius decreases with increasing TH0 energy and the effective pixel area is continuously reduced. The variation in radius depends on the manner of energy deposition in the silicon pixels and can be predicted by Monte-Carlo simulations. Calculations we have performed using the package CASINO [13], show that for 60 and 80 keV electrons the average radii for deposition of the full beam energy are 11 $\mu$m and 20 $\mu$m respectively. Thus, in the limit where the threshold energy TH0=$E_0$, hits are only detected if the electron strikes the pixel at a maximum distance from its centre defined by the pixel half-width (55 $\mu$m / 2) minus the average radius for full energy deposition. This yields values of radius 16.5 $\mu$m (60 keV) or 7.5 $\mu$m (80 keV) from the pixel centre. As these radii refer to circular sub-pixel areas then pixels are reduced to 8.4% and 1.9% of their full area for 60 and 80 keV electrons respectively.

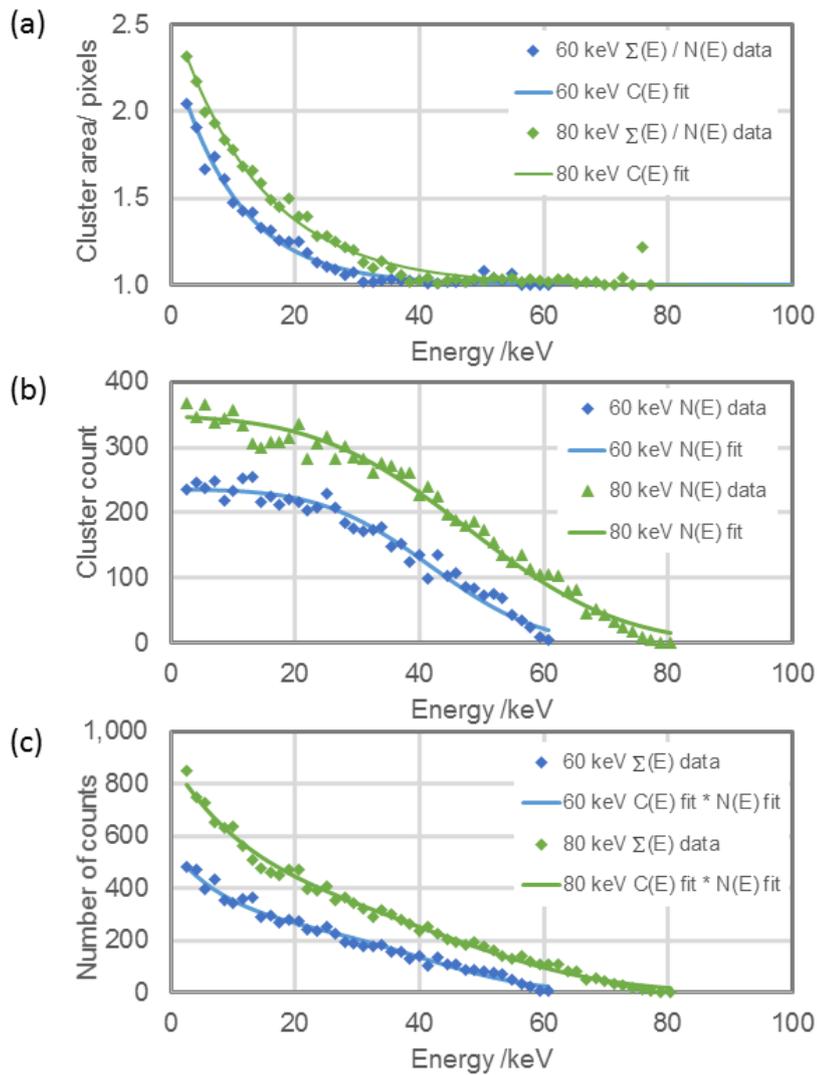

Figure 10. Analysis of $E_0 = 60$ and 80 keV electron cluster data as a function of the threshold energy, TH0. (a) Plot showing $\Sigma(E)$ data / $N(E)$ data and fitted cluster area function $C(E)$. (b) Plot of $N(E)$ data and fit. (c) Plot of $S(E)$ data and fit resulting from the product of $C(E)$ and $N(E)$ fits.

Combining the areal functions C(E) and n(E) it is possible to compute the average response of the detector to electrons impacting at a single point, i.e. the Point Spread Function (PSF). At TH0 values < $E_0/2$, n(E) = 1 and C(E) dictates the average cluster area to be greater than 1 pixel. As TH0 approaches $E_0/2$, C(E) tends to 1 and n(E) then dictates the overall response, possessing values <1, and which tend to zero at TH0 = $E_0$. The radius of the clusters are given by $R = \sqrt{C(E)n(E)}/2$ where R plays the role of the variance in an assumed Gaussian PSF. As n(E) possesses values in the range 1 to 0 with increasing TH0 value, the radius R is smaller than a single pixel. Therefore, an oversampling factor, *M* must be used to properly compute the PSF. The expression for the PSF is given by:

$$PSF(r) = e^{-\frac{1}{\sqrt{2}} \times \left(\frac{r}{RM}\right)^2} \tag{6}$$

where *r* is spatial distance from the pixel centre, *R* is the cluster radius and *M* the oversampling factor. For the PSF computations an oversampling factor *M* = 17 was found to be sufficient. *MTF*s were computed directly from the PSF by Fourier transformation and are shown in figures 11(a) & (b) at 60 keV and 80 keV, respectively for a range of TH0 values.

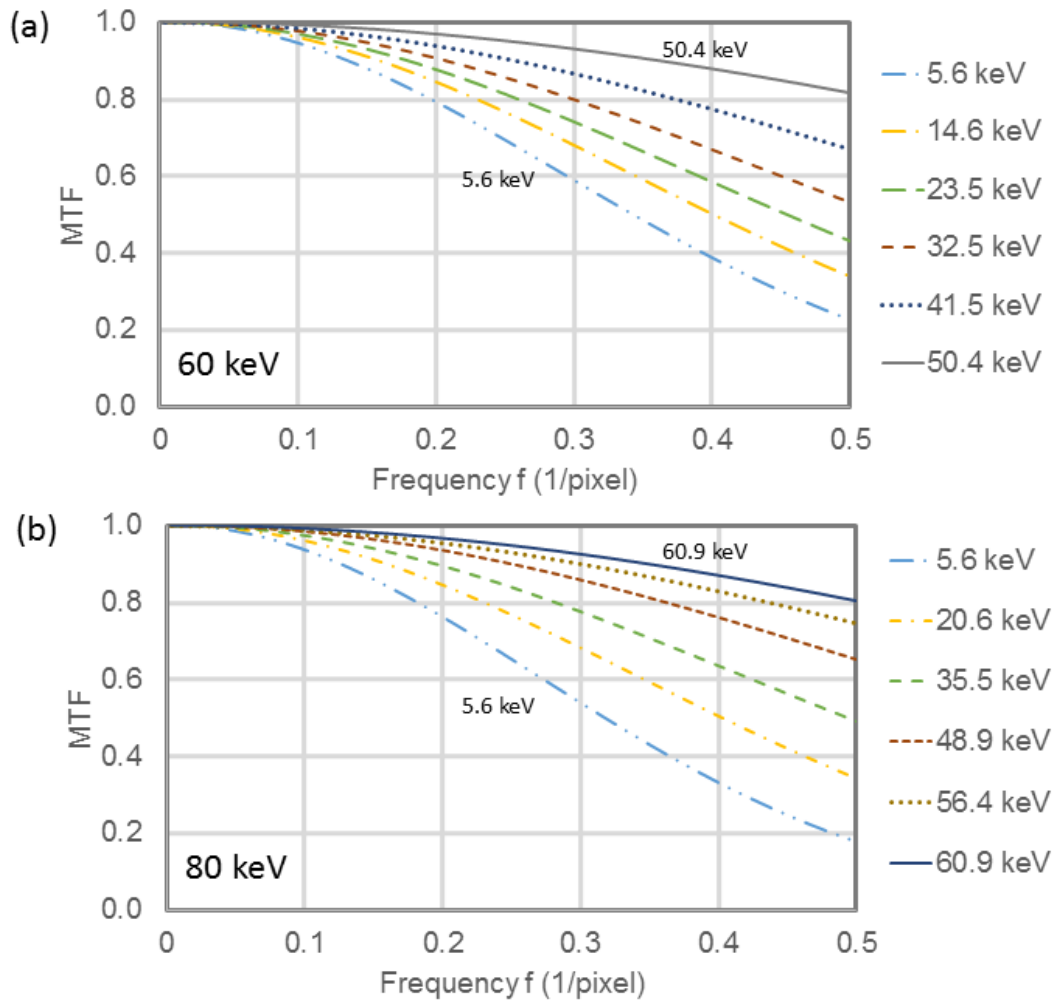

*Figure 11. Calculated MTF curves based on synthesized PSFs from single electron event characterisation. (a) 60keV MTFs. (b) 80keV MTFs.*

How the *MTF* response predicted by the single electron event characterization agrees with values obtained from knife edge measurement is assessed in figures 12 & 13, where the *MTF* values at the Nyquist frequency are plotted as a function of threshold energy. It can be seen that a linear variation of *MTF* at Nyquist with respect to threshold energy is obtained from the single electron event analysis and that the gradients are within 1.7× that for the knife edge derived data. This good agreement supports our understanding of electron energy deposition in the detector and demonstrates the prospect for performance characterization of a counting detector purely by investigating single electron events.

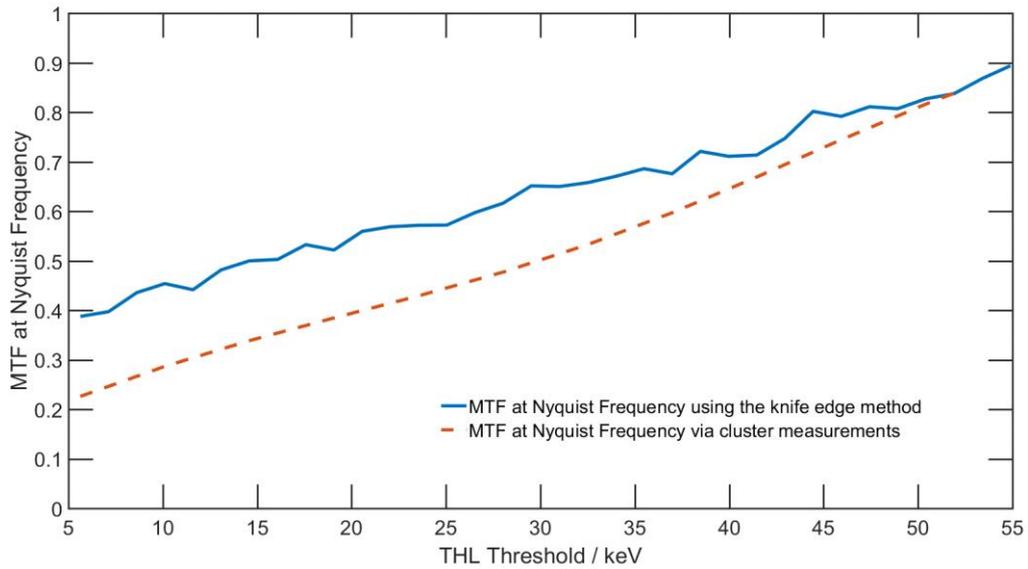

*Figure 12: Comparison of MTF using the established knife edge technique and the cluster counting method. The plot shows MTF at the Nyquist frequency at 60 keV as a function of the energy threshold (TH0) values using SPM.*

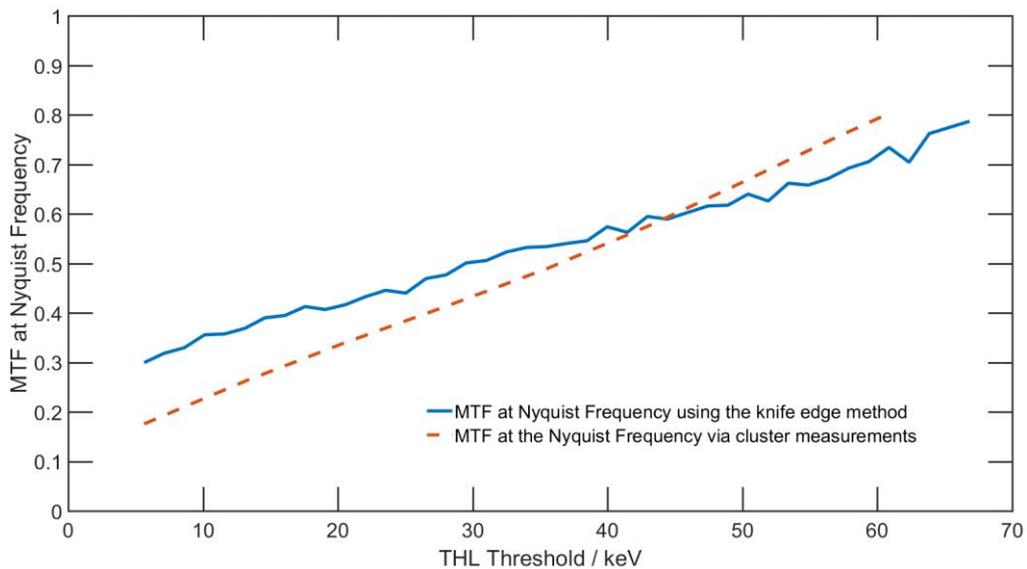

*Figure 13: Comparison of MTF using the established knife edge technique and the cluster counting method. The plot shows MTF at the Nyquist frequency at 80 keV as a function of the lower discriminator threshold (TH0) values using SPM.*

Short, 1μs exposures, also provide insight into the operational performance of CSM. In figures 14 (a) and (b), at 80keV and where TH0 and TH1 are set to energy values just above the detector thermal noise floor, it can be seen that single electron events are indeed recorded as single pixels, as designed. As good as this seems, it does not necessarily deliver ideal detector performance however. Figure 14(c) shows a plot of integrated intensity from longer, 10ms exposure, flat field images for both SPM and CSM as a function of threshold energy value. For the SPM data, the intensity decreases strongly as a function of threshold energy, from 7.9 million counts at TH0=3.0 keV to 0 counts at TH0=80 keV (the beam energy, $E_0$). This is the same response that we have characterized by fitting the S(E)=C(E) × N(E) functions in figures 9 & 10 for single electron events.

Figure 14(c) highlights that CSM operation removes much of the variation in integrated intensity, returning an almost constant number of counts when scanning threshold energy up to 60 keV ($3E_0/4$). Close inspection of the CSM data in figure 14(c) reveals however that there is a small linear decrease in counts from 3.22 million at TH1 = 19.7 keV to 2.74 million at TH1=59.9 keV. This can be attributed to the CSM algorithm not providing perfect correction for ~15% of electron events at this beam energy, most likely returning two separated single pixels. Such events are likely to have been ones in which an incident electron lost significant amounts of energy in spatially separated pixels of two adjacent 2x2 CSM pixel blocks. This could lead to the arbitration circuitry identifying two hits, rather than one. For the CSM data, at TH1 energy threshold values >60 keV in figure 14(c) it can be seen that the integrated intensity suddenly decreases, reaching zero at 80 keV. The width of the transition here relates to the decreasing probability of the CSM algorithm being able to recover all of the deposited charge, as in most cases at least some charge is deposited in an adjacent 2x2 pixel block.

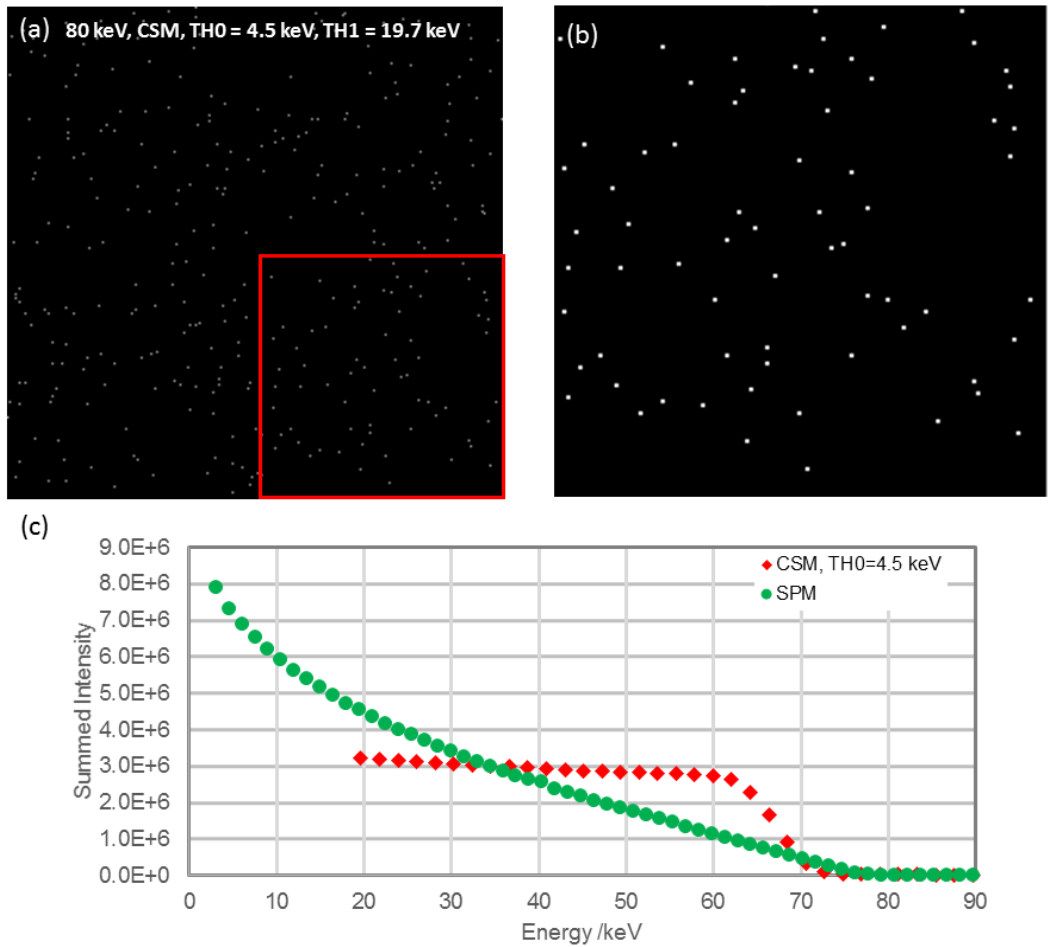

*Figure 14. (a) Single 80keV electron events captured using Charge Summing Mode (CSM) with TH0 = 4.5 keV and TH1 = 19.7 keV and exposure time = 10 $\mu$s. (b) Magnified images of region inside the red box marked in (a) highlighting that electron events are recorded as single pixel events by CSM. (c) Comparison of integrated intensity in flat-field images obtained with exposure time = 10 ms for SPM & CSM with respect to energy threshold.*

# 5. APPLIED IMAGING PERFORMANCE

In this section we demonstrate new imaging capabilities enabled by the pixel architecture of the Medipix3 chip. Each pixel contains two 12 bit counters which offers operational flexibility towards different experimental requirements. For example, it is possible to acquire images with zero gap time between them, enabled by counting into one 12-bit register while simultaneously reading out the other 12-bit register containing counts from the previous image exposure.

It is also possible to configure the two counters as a single 24-bit counter to access a 1 to 16.7 million dynamic range. Such capability directly benefits the quantitative recording of diffraction patterns where the central spot usually has to be blocked with a pointer. Figure 15(a) shows an acquisition of a diffraction pattern of Au nanocrystals on a carbon support (cross-grating replica sample, Agar AGF7016-7) obtained with a parallel beam and current reduced by use of a 10mm condenser aperture in order to ensure the arrival rate <1MHz in the central spot. From an exposure time of 16 seconds the profile in figure 15(b) show the number of counts to vary from a maximum intensity ~10 million counts in the central spot to a minimum intensity ~ 3000 counts at the edge of the pattern.

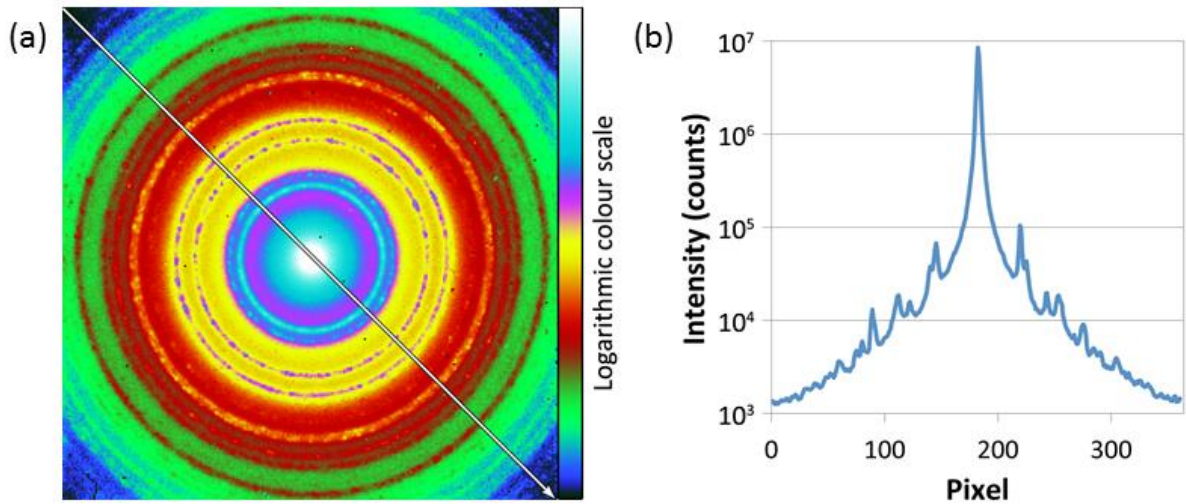

*Figure 15. 24-bit depth acquisition of a diffraction pattern. (a) Acquired diffraction pattern with logarithmic colour scale making visible diffraction features across the full intensity range. (b) Singe-line profile along the pattern diagonal highlighting the dynamic range of the information contained in the pattern.*

Figure 16 demonstrates, directly through images, the variation in *MTF* response of the detector according to threshold energy selection and mode (SPM or CSM) already quantified in figures 2-6. At a microscope indicated magnification of 2 million × the cross-grating replica sample was imaged in TEM mode with a beam energy of 60keV. In figure 16(a), in SPM mode with TH0=20 keV, exposure time = 500ms, Au crystals are easily observed upon the amorphous carbon support. We have not calibrated precisely the magnification obtained for the Medipix3 detector in the top mount position, however, for the purposes of judging approximate scale, the typical sizes of the Au crystals are in the range 5-10 nm. In figure 16(b), the effect of changing the threshold energy to >$E_0/2$, TH0=40 keV, highlighted the improvement in *MTF* performance, with both lattice fringes and Moiré contrast appearing within the Au crystals (regions indicated by the dashed boxes). The lattice fringes themselves are close to the Nyquist frequency, with periodicities of a few pixels. In figure 16(b), selection of the higher threshold energy resulted in 2.1× lower counts and so an increased exposure time of 1000ms was used to maintain the signal to noise ratio. Figure 16(c) shows that by operating in CSM mode, where both TH0 and TH1 were set to values just above the detector thermal noise floor, lattice fringes and Moiré were visible in an image with 500ms exposure time and with mean counts similar to that in figure 16(a). In other words, figure 16(c) demonstrates the provision of simultaneous high *DQE* and *MTF* described in figures 6-8.

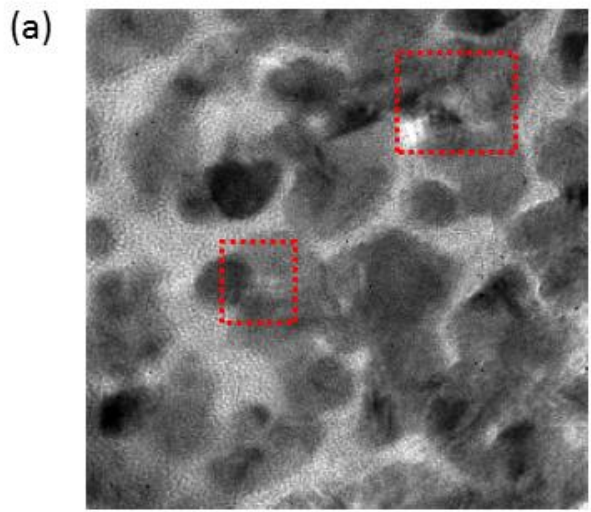

SPM 60 keV, TH0 = 20 keV, 500ms

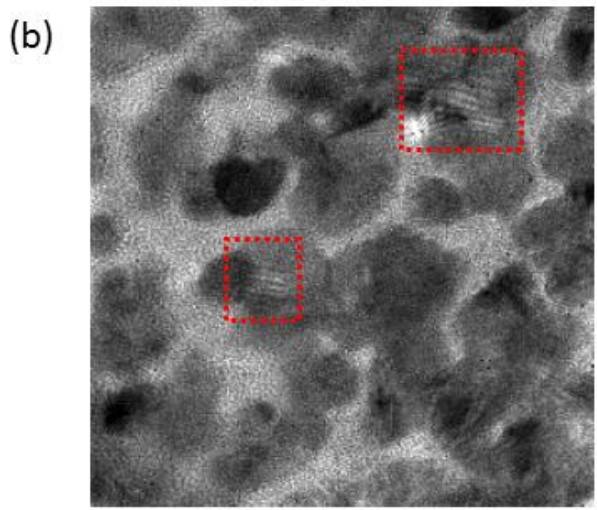

SPM 60 keV, TH0 = 40 keV, 1000ms

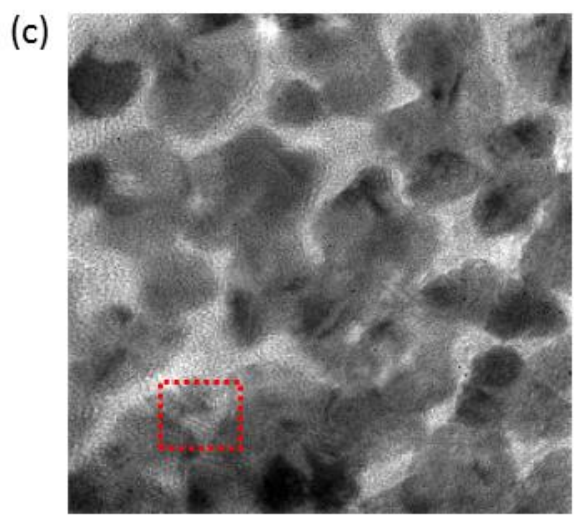

CSM 60 keV, TH0 = 3.0 keV, TH1=19.7 keV, 500ms

*Figure 16 Images of Au nano-crystals of a cross-grating replica sample imaged at 60keV in (a) SPM mode with TH0=20 keV, exposure = 500ms, (b) SPM mode with TH0=40keV, exposure = 1000 ms, (c) CSM mode with TH0=3.0 keV, TH1= 19.7 keV and exposure = 500ms. Red dashed boxes in (a) & (b) refer to identical sample regions and highlight the absence / presence of lattice/Moire fringes. The red dashed box in (c) highlights a different sample region in which lattice fringes are observed.*

## 4. CONCLUSIONS

Across all of our investigations we have performed a comprehensive analysis of the imaging response of the Medipix3 detector at electron beam energies most compatible with the 300μm thick Si sensor. Our measurements of *MTF* and *DQE* in the single pixel mode using conventional knife edge and flat field image methods are in agreement with trends already observed for the Medipix2 detector [4], as expected. Building upon these results, we have gained further insight into the SPM performance through analyzing single electron events and producing an empirical model that can be used to directly predict the *MTF* response of the detector.

Our empirical model agrees well with the accepted knife edge method of measurement and provides insight into the variation of integrated intensity that arises at low threshold energies due to the area of electron hit clusters and at high thresholds due to the reduction in the effective pixel size. The latter phenomenon is responsible for obtaining *MTF* values which exceed the theoretical response of an ideal pixelated detector, but at the expense of operating with a vastly reduced *DQE*. Prediction of *MTF* response from single electron events has been reported previously [15]. Our method, differs in that it synthesizes a PSF based on empirical fitting of integrated intensity and single event counting and is easier to implement across datasets where images can be acquired as a function of detector threshold energy.

We have demonstrated that the Charge Summing Mode (CSM) of the Medipix3 results in significant and simultaneous improvement in *MTF* and *DQE* at both electron beam energies. However, due to the mechanism of energy loss in the sensor material, we have shown that the CSM algorithm does not provide perfect identification of all single electron events or recovery of all spatially distributed charge. These factors most likely explain why the CSM *MTF* and *DQE* responses are excellent but below that of an ideal detector. Thus, it is clear that CSM could present obvious applications for efficient imaging of electron beam sensitive materials.

Beam energies of 60-80 keV are highly relevant for imaging 2D materials such as graphene, in order to prevent kinetic knock-on damage. Beam energies from 120-300keV are much more commonly used in TEMs because they enable higher spatial resolution and imaging of thicker samples. Such beam energies will lead to large spatial dispersion in a silicon sensor material [5] and we propose to study both SPM and CSM operation in order to understand whether the latter algorithm can provide performance improvements in such a regime and whether switching the sensor material to CdTe or GaAs would present similar near ideal levels of detection performance.

## ACKNOWLEDGEMENTS

We gratefully acknowledge the contributions of Dr S. McFadzean and Mr D. Doak who provided expert technical assistance and fabrication of the detector housing.

This work was partly supported by EPSRC grant EP/M009963/1 (Fast Pixel Detectors: a paradigm shift in STEM imaging). Financial support from the European Union under the Seventh Framework Program under a contract for an Integrated Infrastructure Initiative (Ref 312483-ESTEEM2) is gratefully acknowledged.

# REFERENCES

[1]  McMullan G., Faruqi A.R., Henderson R., Guerrini N., Turchetta R., Jacobs A. and Hoften G. van, Experimental observation of the improvement in MTF from backthinning a CMOS direct electron detector, Ultramicroscopy 109 (2009), 401-1147.

[2]  R. Ballabriga, M. Campbell, E. Heijne, X. Llopart, L. Tlustos, W. Wong, Medipix3: A 64 k pixel detector readout chip working in single photon counting mode with improved spectrometric performance, Nuclear Instruments and Methods in Physics Research Section A: Accelerators, Spectrometers, Detectors and Associated Equipment, Volume 633, Supplement 1, May 2011, Pages S15-S18, ISSN 0168-9002, http://dx.doi.org/10.1016/j.nima.2010.06.108. (http://www.sciencedirect.com/science/article/pii/S0168900210012982).

[3]  Tate M.W., Purohit P., Chamberlain D., Nguyen K.X., Hovden R., Chang C.S., Deb P., Turgut E., Heron J.T., Schlom D.G., Ralph D.C., Fuchs G.D., Shanks K.S., Philipp H.T., Muller D.A and Gruner SM, High Dynamic Range Pixel Array Detector for Scanning Transmission Electron Microscopy, Microsc. Micronal 1 (2016) 237-49.

[4]  McMullan G., Cattermole D.M., Chen S., Henderson R., Llopart X., Summerfield C., Tlustos L. and Faruqi A.R., Electron imaging with Medipix2 hybrid pixel detector, Ultramicroscopy 107 (2007), 401-413

[5]  McMullan G. et al., Detective Quantum Efficiency of electron area detectors in electron microscopy, Ultramicroscopy 109 (2009), 1126-1143.

[6]  R. Ballabriga, *The Design and Implementation in 0.13 $\mu$m CMOS of an Algorithm Permitting Spectroscopic Imaging with High Spatial Resolution for Hybrid Pixel Detectors.* PhD Thesis. Universitat Ramon Llull (2009). Available at *http:// cds.cern.ch/record/1259673/files/CERN-THESIS-2010-055.pdf*

[7] Pennicard D., Ballabriga R., Llopart X., Campbell M. and Graafsma H., Simulations of charge summing and threshold dispersion effects in medipix3, Nucl. Instrum. & Meth. In Physics Research A 636 (2011) 74-81.

[8] Meyer R.R., Kirkland A.I., Dunin-Borkowski R.E. and Hutchison J.L., Experimental characterisation of CCD cameras for HREM at 300 kV, Ultramicroscopy 85,1 (2000) 9-13.

[9] S. McVitie, D. McGrouther, S. McFadzean, D.A. MacLaren, K.J. O'Shea, M.J. Benitez, Ultramicroscopy 152, 57 (2015).

[10] Plackett R. and Omar D., Merlin: Medipix3RX Quad Readout – Manual and User Guide, Oxfordshire: DIAMOND LIGHT SOURCE LIMITED, 2015.

[11] Plackett R., Horsewell I., Gimenez E.N., Marchal J.,Omar D., Tartoni N., Merlin: a fast versatile readout system for Medipix3, Journal of Instrumentation, Vol. 8, Article Number C01038 (2013).

[12] Ruskin R.S., Zhiheng Yu and Nikolaus Grigorieff., Quantitative characterization of electron detectors for transmission electron microscopy, Journal of Structural Biology 184 (2013) 385-393.

[13] Drouin D., Couture A.L., Joly D., Xavier T., Vincent A. and Raynland G., CASINO V2.42-A Fast and Easy-to-use Modelling Tool for Scanning Electron Microscopy and Microanalysis users, SCANNING Vol. 29, Issue 3 (2007) 92-101. (see also http://www.gel.usherbrooke.ca/casino/links.html).

[14] R Ghadimi, I Daberkow, C Kofler, P Sparlinek and H Tietz (2011). Characterization of 16 MegaPixel CMOS Detector for TEM by Evaluating Single Events of Primary Electrons. Microscopy and Microanalysis, 17 (Suppl. 2) , pp 1208-1209. doi:10.1017/S143192761100691X.